\documentclass{ws-procs9x6-cpt25}
\makeatletter

\usepackage{booktabs}

\begin{document}

\newcommand{\refeq}[1]{(\ref{#1})}
\def\etal {{\it et al.}}
\def\PRL {{Phys.\ Rev.\ Lett.}}
\def\PRA {{Phys.\ Rev.\ A}}
\def\PRC {{Phys.\ Rev.\ C}}
\def\PRD {{Phys.\ Rev.\ D}}
\def\PLA {{Phys.\ Lett.\ A}}
\def\EPL {{Europhys. Lett.}}
\def\PPNP {{Prog.\ Part.\ Nucl.\ Phys.} }
\def\NIMA {{Nucl.\ Inst.\ Meth.\ A} }
\def\NIM {{Nucl.\ Inst.\ Meth.} }
\def\EPJA {{Eur.\ Phys.\ J.\ A}} 
\def\EPJB {{Eur.\ Phys.\ J.\ B}} 
\def\EPJC {{Eur.\ Phys.\ J.\ C}} 
\def\EPJD {{Eur.\ Phys.\ J.\ D}}


\title{Neutron EDM Experiment\\ with an Advanced Ultracold Neutron Source at TRIUMF}

\author{
	\centering 
	T.\ Higuchi,$^{1,2}$ 
	B.\ Algohi,$^3$ 
	D.\ Anthony,$^4$ 
	L.\ Barr\'on-Palos,$^5$ 
	M.\ Bradley,$^6$ 
	A.\ Brossard,$^4$ 
	T.\ Bui,$^3$ 
	J.\ Chak,$^4$ 
	R.\ Chiba,$^{7,4}$
	C.\ Davis,$^4$ 
	R.\ de Vries,$^8$ \\
	K.\ Drury,$^4$ 
	D.\ Fujimoto,$^4$
	R.\ Fujitani,$^{9,1}$ 
	M.\ Gericke,$^3$ 
	P.\ Giampa,$^4$ \\
	R.\ Golub,$^{10}$
	T.\ Hepworth,$^8$
	G.\ Ichikawa,$^{11}$ 
	S.\ Imajo,$^2$ 
	A.\ Jaison,$^3$ \\
	B.\ Jamieson,$^8$ 
	M.\ Katotoka,$^8$
	S.\ Kawasaki,$^{12}$
	M.\ Kitaguchi,$^{13}$\\
	W.\ Klassen,$^{14}$
	E.\ Korkmaz,$^{15}$
	E.\ Korobkina,$^{10}$
	M.\ Lavvaf,$^3$
	T.\ Lindner,$^{4,8}$
	N.\ Lo,$^4$\\
	S.\ Longo,$^3$
	K.\ Madison,$^{14}$
	Y.\ Makida,$^{12}$
	J.\ Malcolm,$^4$
	J.\ Mammei,$^3$\\
	R.\ Mammei,$^8$
	Z.\ Mao,$^{14}$
	C.\ Marshall,$^4$
	J.W.\ Martin,$^8$
	M.\ McCrea,$^8$\\
	E.\ Miller,$^{14}$
	M.\ Miller,$^{16}$
	K.\ Mishima,$^{2}$
	T.\ Mohammadi,$^3$
	T.\ Momose,$^{14}$\\
	T.\ Okamura,$^{12}$
	H.J.\ Ong,$^2$
	R.\ Patni,$^4$
	R.\ Picker,$^{4,7}$
	K.\ Qiao,$^{17,2}$\\
	W.D.\ Ramsay,$^4$
	W.\ Rathnakela,$^3$
	D.\ Salazar,$^{7,4}$
	J.\ Sato,$^{12}$
	W.\ Schreyer,$^{4,18}$
	T.\ Shima,$^2$
	H.M.\ Shimizu,$^{12}$
	S.\ Sidhu,$^4$
	S.\ Stargardter,$^3$
	P.\ Switzer,$^8$\\
	I.\ Tanihata,$^2$
	Tushar,$^3$
	S.\ Vanbergen,$^{4}$
	W.T.H.\ van\ Oers,$^{3,4}$\\
	Y.\ Watanabe,$^{12}$
	N.\ Yazdandoost,${^4}$
	Q.\ Ye,$^{14}$
	A.\ Zahra,$^3$
	and M.\ Zhao$^4$\\ \vspace{.2
		em}
	(TUCAN Collaboration)\\
}

{
	\vspace{1em}
	\address{$^1$Institute for Integrated Radiation and Nuclear Science, Kyoto University,\\ Kumatori, 590-0494, Japan}
	\address{$^2$Research Center for Nuclear Physics, the University of Osaka,\\ Ibaraki, 567-0047, Japan}
	\address{$^3$The University of Manitoba, Winnipeg, MB R3T 2N2, Canada}
	\address{$^4$TRIUMF, Vancouver, BC V6T 2A3, Canada}
	\newpage
	\address{$^5$
		Instituto de F\'isica, Universidad Nacional Aut\'onoma de M\'exico,\\ Apartado Postal 20-364, 01000, Mexico}
	\address{$^6$The University of Saskatchewan, Saskatoon, SK S7N 5A2, Canada}
	\address{$^{7}$Simon Fraser University, Burnaby, BC V5A 1S6, Canada}
	\address{$^8$The University of Winnipeg, Winnipeg, MB R3B 2E9, Canada}
	\address{$^9$Department of Nuclear Engineering, Kyoto University, Kyoto, 615-8540, Japan}
	\address{$^{10}$North Carolina State University, Raleigh, NC 27695-8202, USA}
	\address{$^{11}$Institute of Materials Structure Science,\\ High Energy Accelerator Research Organization (KEK), Tokai,  319-1106, Japan}
	\address{$^{12}$Institute for Particle and Nuclear Studies,\\ High Energy Accelerator Research Organization (KEK), Tsukuba, 305-0801, Japan}
	\address{$^{13}$Nagoya University, Nagoya, 464-8601, Japan}
	\address{$^{14}$The University of British Columbia, Vancouver, BC V6T 1Z4, Canada}
	\address{$^{15}$The University of Northern British Columbia, Prince George, BC V2N 4Z9, Canada}
	\address{$^{16}$McGill University, Montr\'eal, QC H3A 0G4, Canada}
	\address{$^{17}$Graduate School of Science, the University of Osaka, Toyonaka, 560-0043, Japan}
	\address{$^{18}$Oak Ridge National Laboratory, Knoxville, TN 37830, USA}
	
}

\begin{abstract}
	The \underline{T}RIUMF \underline{U}ltra\underline{c}old \underline{A}dvanced \underline{N}eutron (TUCAN) collaboration has been developing a high-intensity ultracold neutron (UCN) source aimed at searching for the neutron electric dipole moment (EDM) with a sensitivity goal of $10^{-27}\,e\,\mathrm{cm}$. This article reports on recent progress in the commissioning of the UCN source and the development of the neutron EDM spectrometer. In its final configuration, the accelerator-driven super-thermal UCN source will enable a neutron EDM experiment with two orders of magnitude improved statistics compared to the current best experiment. Substantial progress in 2024 allowed the collaboration to operate the complete source system, with the exception of the liquid deuterium cold moderator, resulting in the first production of UCNs. The status of the EDM spectrometer is also presented, with emphasis on UCN handling components and magnetic subsystems relevant to field control, shielding, and magnetometry.
\end{abstract}

\bodymatter

\section{Background}
Measurements of the neutron electric dipole moment (EDM) occupy an important place in today's particle physics. A finite neutron EDM would violate time-reversal symmetry, which, on the basis of  CPT symmetry, would imply CP violation. Experiments so far have found results consistent with zero, placing stringent constraints on the QCD $\bar{\theta}$ term and on theories beyond the Standard Model, such as supersymmetric extensions or multi-Higgs-doublet models.\cite{Pospelov2005,Engel2013,Liang2023,Hou2024,Abe2018}

State-of-the-art neutron EDM experiments utilize stored ultracold neutrons (UCNs) to measure the EDM by observing the precession of the neutron spin under electric and magnetic fields. UCNs are neutrons with kinetic energies of around $300$~neV or less, which can be stored in containers coated with materials of high Fermi potential. This enables neutron EDM experiments with reduced systematic effects related to neutron velocity and with observation times on the order of 100~s.
The current best upper limit, obtained at the Paul Scherrer Institute, is $|d_n| < 1.8 \times 10^{-26}\,e\mathrm{cm}$ (90\% C.L.).\cite{Abel2020} This result is limited by the statistical uncertainty arising from the number of UCNs available during the experiment.

To improve upon this limit and measure the neutron EDM with unprecedented precision, the \underline{T}RIUMF \underline{U}ltra\underline{c}old \underline{A}dvanced \underline{N}eutron (TUCAN) collaboration has been developing a high-intensity UCN source, expected to increase the number of usable UCNs by two orders of magnitude. Our goal sensitivity is $10^{-27}\,e\mathrm{cm}$ and is competitive with other next-generation EDM experiments that are in preparation in locations around the world.\cite{Wurm2019,Ayres2021,Wong2023}

\section{Advanced ultracold neutron source at TRIUMF}
The TUCAN source is based on a combination of accelerator-driven spallation neutron production and the super-thermal method with superfluid helium (He-II) as the UCN converter.\cite{Golub1977} High-energy neutrons produced by spallation are decelerated by cold moderators and reach the He-II converter, where they are downscattered to become UCNs. This scheme was developed at Research Center for Nuclear Physics in Japan and has been demonstrated with a prototype UCN source.\cite{Masuda2012} The prototype source was transported to TRIUMF, installed on a dedicated 480-MeV proton beamline driven by the TRIUMF cyclotron,\cite{Ahmed2019a} and was used to produce the first UCNs at TRIUMF in 2017.\cite{Ahmed2019} It was operated until 2019 for characterization of the source itself and tests of UCN handling components.\cite{Ahmed2019,Hansen-Romu2023,Akatsuka2023}

Based on experience with the prototype source, the collaboration has been developing a new, upgraded UCN source featuring a liquid deuterium (LD$_2$) cold moderator optimized for UCN production, a 27-L He-II converter, and a helium cryostat capable of maintaining He-II temperatures near 1~K under the expected 10~W heat load during beam irradiation.\cite{Schreyer2020,Kawasaki2020} The expected UCN yield with a 40~$\mu$A proton beam is $1.4 \times 10^7$~UCN/s,\cite{Schreyer2020} representing a 500-fold increase over the $2 \times 10^4$~UCN/s achieved with the prototype source operated at the nominal 1~$\mu$A beam current.\cite{Ahmed2019}

According to Monte-Carlo UCN transport simulations, the source will enable an EDM experiment with $10^6$~UCNs detected per measurement cycle. This corresponds to a statistical uncertainty of $10^{-25}\,e\mathrm{cm}$ per cycle and will allow us to reach the target sensitivity of $10^{-27}\,e\mathrm{cm}$ within 280 days of data taking.\cite{Sidhu2023}

\section{Recent status}
\paragraph{Status of the UCN source}
The year 2024 saw major progress in the manufacturing of components for the TUCAN source. In April, fabrication of the vessels for the UCN production volume was completed. This was followed by the completion of key components, including the main heat exchanger of the helium cryostat and the UCN guides for extracting the produced UCNs from the source. By the end of October, all hardware for the TUCAN source, except the LD$_2$ system, was in place. The helium cryostat, built in Japan and shipped to TRIUMF in 2021, has undergone a series of cooldowns to test its performance, initially using natural helium and later using dedicated helium isotopes: $^3$He for the cryostat cooling line and isotopically pure $^4$He for the UCN converter.

In November 2024, the full TUCAN source system apart from the LD$_2$ moderator was commissioned. The UCN production volume was filled with isotopically pure $^4$He cooled to superfluid temperatures.
Although the absence of the LD$_2$ moderator was expected to significantly reduce the yield, a production rate of $\sim 10^4$~UCN/$\mu$A was still anticipated. However, no significant UCN events above background were observed. Air contamination in the isotopically pure $^4$He batch used in the production volume was strongly suspected as the cause, supported by several observations made before and during the run.
Nevertheless, the cryogenic performance of the source was successfully characterized using both a heater and the proton beam, demonstrating sufficient cooling capacity for operation at 40~$\mu$A.\cite{Martin2025} Cold neutron fluxes from the D$_2$O moderator were measured by the gold foil activation method to benchmark Monte Carlo N-Particle (MCNP) simulations used in the source design.\cite{Schreyer2020}

After a large amount of contaminants had been removed from the $^4$He, another attempt was made in June 2025 with a custom-built $^3$He/$^4$He purification system installed on the isotopically pure $^4$He inlet. This led to the first detection of UCNs extracted from the TUCAN source. The initial results indicate good agreement between the estimated and observed UCN yields.\cite{Algohi2025}

\paragraph{Development of the EDM spectrometer}
In parallel with the development of the UCN source, subsystems of the neutron EDM spectrometer have been advanced. Tests of key UCN handling components have been conducted using the pulsed UCN source at J-PARC/MLF,\cite{Imajo2016} which has supported various experiments, including transmission through UCN guides, storage in a prototype EDM measurement cell, and UCN polarization using magnetized iron films.\cite{Imajo2023,Vanbergen2025,Higuchi2024} The cold-neutron beamline MINE2 at JRR-3 has also been used for complementary characterization of polarization films and surface coatings relevant to UCN applications.\cite{Hino2021,Higuchi2024}

One crucial aspect of the neutron EDM experiment is the precise control of magnetic fields, an area in which active development is ongoing. In the final neutron EDM experiment, the effective stability required for the spin-holding magnetic field ($B_0 \approx 1~\mu$T) is about 10~fT over averaging times on the order of 100~s. This is achieved through the use of advanced magnetic shielding and an optically pumped $^{199}$Hg magnetometer that monitors magnetic field variations in the same volume occupied by UCNs.

A multilayer magnetically shielded room (MSR), consisting of five mu-metal layers and one copper layer, has recently been constructed. It has an outer dimension of 3.5~m and an inner room measuring 2.25~m on each side.
The shielding performance was characterized at each stage of assembly, demonstrating shielding sufficient to achieve field stability at the 10~pT level. During the tests, ambient magnetic fields from the TRIUMF cyclotron, reaching up to 370~$\mu$T around the MSR, were found to degrade the shielding performance. Compensation coils have been designed to address this issue and are currently under construction.\cite{Higuchi2022}

The $^{199}$Hg co-magnetometer under development at the University of British Columbia has demonstrated a sensitivity on the order of 100~pT, currently limited by a few aspects of the prototype setup, such as magnetic field inhomogeneities and wall collisions in the small test cell.
Once integrated into the full-size EDM measurement cell in the MSR, it is expected to reach the sensitivity required to monitor magnetic field variations at the 10~fT level.

Generation of a uniform $B_0$ field is also important, since magnetic-field inhomogeneities are one of the dominant sources of systematic uncertainty.\cite{Abel2020,Abel2019} To map magnetic field distributions inside the MSR, an array of optically pumped Cs atomic magnetometers have been developed. Recent experiments in a test setup demonstrated a stability of 90~fT over an averaging time of 150~s.\cite{Klassen2024} The system supports operation of multiple sensors using a single laser diode; it has been successfully tested with 3 sensors and is being prepared for deployment with 20.\cite{Klassen2024} Coil systems with different purposes, including a self-shielded $B_0$ coil and a set of shim coils to correct for magnetic field inhomogeneities, have been designed and are currently under construction.\cite{McCrea2022}

\section{Potential tests of Lorentz symmetry with TUCAN}
The advanced magnetic subsystems developed for the neutron EDM experiment can, in principle, be applied to clock-comparison tests of Lorentz symmetry, in which Lorentz-violating effects would manifest as sidereal modulations in the spin-precession frequencies of magnetometers. Here, the use of multiple spin species is essential to cancel background magnetic-field drifts and isolate components that would couple differently to each.
Physical systems accessible with TUCAN include neutron/$^{199}$Hg, Cs/$^{199}$Hg, and $^{199}$Hg/$^{201}$Hg. Since the Cs magnetometer measures the precession of an electron spin and the Hg magnetometer measures that of a nuclear spin, the Cs/$^{199}$Hg system is sensitive to Lorentz-violating effects in the electron, proton, and neutron sectors.\cite{Hunter1999,Berglund1995,Peck2012} In contrast, the neutron/$^{199}$Hg and $^{199}$Hg/$^{201}$Hg systems are primarily sensitive to the neutron sector.\cite{Altarev2011,Lamoreaux1986,Stadnik2015}
Constraints on minimal Standard-Model Extension (SME) coefficients obtained from previous experiments using such systems are excerpted from the data tables and summarized in Table~\ref{tab:Lorentz}.\cite{datatables} A recently developed theoretical treatment further allows constraints to be placed on nonminimal Lorentz-violating coefficients from these experiments.\cite{Kostelecky2018a}

Some clock-comparison experiments, including that in Ref.~\refcite{Peck2012}, employed rotating apparatus to enhance sensitivity. While this approach is not feasible with the TUCAN MSR, the projected effective magnetic field stability of 10--100~fT per 100-s cycle makes it possible to improve the limits on these coefficients. For reference, upper limits on Lorentz-violating signals in units of magnetic fields were 14~fT in Ref.~\refcite{Berglund1995} and 4~fT in Ref.~\refcite{Peck2012}.

\begin{table}[t]
	\tbl{List of minimal SME coefficients accessible to the TUCAN EDM spectrometer, excerpted from Ref.\ \protect\refcite{datatables}. For the limit from neutron/Hg systems, the result from Ref.\ \protect\refcite{Altarev2011} is taken, which updates the earlier result in Ref.\ \protect\refcite{Altarev2009}.}
	{
		\begin{tabular}{@{}ccccc@{}}\toprule
			Sector  & Coefficients & Limits (GeV) & Systems & Refs. \\ \colrule
			$e$ & $r_e$ & $< 3.2\times 10^{-24}$  & Hg/Cs & [\refcite{Hunter1999}] \\
			& $|\tilde{b}_J|,\,(J=X,Y)$ & $< 10^{-27}$  & Hg/Cs & [\refcite{Berglund1995,Kostelecky1999}]\\
			& $|\tilde{d}_J|,\,(J=X,Y)$ & $< 10^{-22}$  & Hg/Cs & [\refcite{Berglund1995,Kostelecky1999}] \\
			& $|\tilde{g}_{D,J}|,\,(J=X,Y)$ & $< 10^{-22}$  & Hg/Cs & [\refcite{Berglund1995,Kostelecky1999}] \\
			\midrule
			$p$ & $\tilde{b}_Z$ & $< 7 \times 10^{-29}$  & Hg/Cs & [\refcite{Peck2012}] \\
			& $\tilde{b}_\perp$ & $< 4 \times 10^{-30}$  & Hg/Cs & [\refcite{Peck2012}] \\
			& $|\tilde{b}_J|,\,(J=X,Y)$ & $< 10^{-27}$  & Hg/Cs & [\refcite{Berglund1995,Kostelecky1999}] \\
			& $\tilde{d}_J,\tilde{g}_{D,J},\,(J=X,Y)$ & $< 3\times 10^{-28}$  & Hg/Cs & [\refcite{Peck2012}] \\
			& $|\tilde{d}_J|, |\tilde{g}_{D,J}|,\,(J=X,Y)$ & $< 10^{-25}$  & Hg/Cs & [\refcite{Berglund1995,Kostelecky1999}] \\
			\midrule
			$n$ & $\tilde{b}_Z$ & $< 7\times 10^{-30}$  & Hg/Cs & [\refcite{Peck2012}] \\
			& $\tilde{b}_\perp$ & $< 3.7\times 10^{-31}$  & Hg/Cs & [\refcite{Peck2012}] \\
			& $b_\perp$ & $< 1\times 10^{-29}$  & $n$/Hg & [\refcite{Altarev2011}] \\
			& $r_n$ & $< 1.5\times 10^{-30}$  & Hg/Cs & [\refcite{Hunter1999}]\\
			& $|\tilde{b}_J|,\,(J=X,Y)$ & $< 10^{-30}$  & Hg/Cs & [\refcite{Berglund1995,Kostelecky1999}] \\
			& $|\tilde{c}_-|, |\tilde{c}_Z|$ & $< 10^{-27}$  & Hg/Hg, Ne/He & [\refcite{Lamoreaux1986,Chupp1989}] \\
			& $\tilde{d}_J,\tilde{g}_{D,J},\,(J=X,Y)$ & $< 3 \times 10^{-29}$  & Hg/Cs & [\refcite{Peck2012}] \\
			& $|\tilde{d}_J|, |\tilde{g}_{D,J}|,\,(J=X,Y)$ & $< 10^{-28}$  & Hg/Cs & [\refcite{Berglund1995,Kostelecky1999}] \\
			\botrule
		\end{tabular}
	}
	\label{tab:Lorentz}
\end{table}

\section{Summary and Outlook}
In this contribution, we report on the recent commissioning progress of the TUCAN source at TRIUMF. We also outline the development status of the EDM spectrometer, with a focus on UCN handling components and the magnetic subsystems essential to the EDM experiment.

The major milestone of the first UCN production from the TUCAN source was achieved in June 2025.
The next significant step is the commissioning of the full source, including the LD$_2$ moderator. The LD$_2$ cryostat was completed in April 2025 and delivered to TRIUMF in May 2025. Integration of the LD$_2$ moderator is expected to enhance the UCN production rate by a factor of 30, unlocking the full performance of the TUCAN source.

In 2026, a site-wide shutdown of the TRIUMF accelerators is scheduled. The TUCAN collaboration will use this period to characterize the performance of spectrometer subsystems and commission the EDM spectrometer in preparation for UCN experiments beginning in 2027.

\section*{Acknowledgments}
We gratefully acknowledge the support of the Canada Foundation for Innovation; the Canada Research Chairs program; the Natural Sciences and Engineering Research Council of Canada (NSERC) SAPPJ-2016-00024, SAPPJ-2019-00031, and SAPPJ-2023-00029; JSPS KAKENHI (Grant Nos.\ 18H05230, 19K23442, 20KK0069, 20K14487, 22H01236, 25H00652 and 25H02182); JSPS Bilateral Program (Grant No.\ JSPSBP120239940); JST FOREST Program (Grant No.\ JPMJFR2237); International Joint Research Promotion Program of Osaka University; RCNP COREnet; the Yamada Science Foundation; the Murata Science Foundation; Yazaki Memorial Foundation for Science and Technology; the Grant for Overseas Research by the Division of Graduate Studies (DoGS) of Kyoto University; the United States Department of Energy (DoE) (grant No.\ DE-FG02-97ER41042); the Universidad Nacional Aut\'onoma de M\'exico--DGAPA program PASPA and grant PAPIIT AG102023.


\begin{thebibliography}{xx}
	
	\bibitem{Pospelov2005}
	M.\ Pospelov and A.\ Ritz, Ann.\ Phys.\ {\bf 318}, 119 (2005).
	
	\bibitem{Engel2013}
	J.\ Engel, M.J.\ Ramsey-Musolf, and U.\ Van\ Kolck, \PPNP\ {\bf 71}, 21 (2013).
	
	\bibitem{Liang2023}
	J.\ Liang \etal\ ($\chi$QCD collaboration), \PRD\ {\bf 108}, 094512 (2023).
	
	\bibitem{Hou2024}
	W.-S.\ Hou, G.\ Kumar, and S.\ Teunissen, \PRD\ {\bf 109}, L011703 (2024).
	
	\bibitem{Abe2018}
	T.\ Abe \etal, \PRD\ {\bf 98}, 075029 (2018).
	
	\bibitem{Abel2020}
	C.\ Abel \etal, \PRL\ {\bf 124}, 081803 (2020).
	
	\bibitem{Wurm2019}
	D.\ Wurm \etal, EPJ Web Conf.\ {\bf 219}, 02006 (2019).
	
	\bibitem{Ayres2021}
	N.J.\ Ayres \etal, \EPJC\ {\bf 81}, 512 (2021).
	
	\bibitem{Wong2023}
	D.K.-T.\ Wong \etal, \NIMA\ {\bf 1050}, 168105 (2023).
	
	\bibitem{Golub1977}
	R.\ Golub and S.K.\ Lamoreaux, \PLA\ {\bf 62}, 338 (1977).
	
	\bibitem{Masuda2012}
	Y.\ Masuda \etal, Phys.\ Rev.\ Lett.\ {\bf 108}, 134801 (2012).
	
	\bibitem{Ahmed2019a}
	S.\ Ahmed \etal, \NIMA\ {\bf 927}, 101 (2019).
	
	\bibitem{Ahmed2019}
	S.\ Ahmed \etal\ (TUCAN Collaboration), Phys.\ Rev.\ C {\bf 99}, 025503 (2019).
	
	\bibitem{Hansen-Romu2023}
	S.\ Hansen-Romu, Ph.D.\ thesis, the University of Manitoba (2023).
	
	\bibitem{Akatsuka2023}
	H.\ Akatsuka \etal, \NIMA\ {\bf 1049}, 168106 (2023).
	
	\bibitem{Schreyer2020}
	W.\ Schreyer \etal, \NIMA\ {\bf 959}, 163525 (2020).
	
	\bibitem{Algohi2025}
	B.\ Algohi \etal, arXiv:2509.02916.
	
	\bibitem{Kawasaki2020}
	S.\ Kawasaki, T.\ Okamura, and the TUCAN Collaboration, IOP Conf.\ Ser.: Mater.\ Sci.\ Eng.\ {\bf 755}, 012140 (2020).
	
	\bibitem{Sidhu2023}
	S.\ Sidhu \etal, EPJ Web Conf.\ {\bf 282}, 01015 (2023).
	
	\bibitem{Martin2025}
	J.W.\ Martin \etal, EPJ Web Conf.\ {\bf 333}, 03004 (2025).
	
	\bibitem{Imajo2016}
	S.\ Imajo \etal, Prog.\ Theor.\ Exp.\ Phys.\ {\bf 2016}, 013C02 (2016).
	
	\bibitem{Imajo2023}
	S.\ Imajo \etal, \PRC\ {\bf 108}, 034605 (2023).
	
	\bibitem{Vanbergen2025}
	S.\ Vanbergen, Ph.D.\ thesis, the University of British Columbia (2025).
	
	\bibitem{Higuchi2024}
	T.\ Higuchi \etal, J.\ Phys.\ Soc.\ Jpn.\ {\bf 93}, 091009 (2024).
	
	\bibitem{Hino2021}
	M.\ Hino and T.\ Oda, Hamon\ {\bf 31}, 36 (2021).
	
	\bibitem{Higuchi2022}
	T.\ Higuchi for the TUCAN Collaboration, EPJ Web Conf.\ {\bf 262}, 01015 (2022).
	
	\bibitem{Abel2019}
	C.\ Abel \etal, \PRA\ {\bf 99}, 042112 (2019).
	
	\bibitem{Klassen2024}
	W.\ Klassen \etal, \EPJC\ {\bf 84}, 1181 (2024).
	
	\bibitem{McCrea2022}
	M.\ McCrea for the TUCAN Collaboration, Proceedings of Science, PoS(PANIC2021), 459 (2022).
	
	\bibitem{Hunter1999}
	L.R.\ Hunter \etal, Proceedings of the Meeting on CPT and Lorentz Symmetry, World Scientific, Singapore, 1999.
	
	\bibitem{Berglund1995}
	C.J.\ Berglund \etal, \PRL\ {\bf 75}, 1879 (1995).
	
	\bibitem{Kostelecky1999}
	V.A.\ Kosteleck\'y and C.D.\ Lane, \PRD\ {\bf 60}, 116010 (1999).
	
	\bibitem{Peck2012}
	S.K.\ Peck \etal, \PRA\ {\bf 86}, 012109 (2012).
	
	\bibitem{Altarev2009}
	I.\ Altarev \etal, \PRL\ {\bf 103}, 081602 (2009).
	
	\bibitem{Altarev2011}
	I.\ Altarev \etal, Physica B {\bf 406}, 2365 (2011).
	
	\bibitem{Lamoreaux1986}
	S.K.\ Lamoreaux \etal, \PRL\ {\bf 57}, 3125 (1986).
	
	\bibitem{Stadnik2015}
	Y.V.\ Stadnik and V.V.\ Flambaum, \EPJC\ {\bf 75}, 110 (2015).
	
	\bibitem{Chupp1989}
	T.E.\ Chupp \etal, \PRL\ {\bf 63}, 1541 (1989).
	
	\bibitem{datatables}
	{\it Data Tables for Lorentz and CPT Violation,}
	V.A.\ Kosteleck\'y and N.\ Russell,
	2025 edition,
	arXiv:0801.0287v18.
	
	\bibitem{Kostelecky2018a}
	V.A.\ Kosteleck\'y and A.J.\ Vargas, \PRD\ {\bf 98}, 036003 (2018).
	
\end{thebibliography}
\end{document}